# Challenges and opportunities in orbitronics

Shunsuke Fukami[1,2], Kyung-Jin Lee[3], and Mathias Kläui[4,5]

1. Research Institute of Electrical Communication, Tohoku University, Sendai 980-8577, Japan
2. Center for Science and Innovation in Spintronics, Tohoku University, Sendai 980-8577, Japan
3. Department of Physics, Korea Advanced Institute of Science and Technology (KAIST), Daejeon, Republic of Korea
4. Institut für Physik, Johannes Gutenberg-Universität Mainz, 55099, Mainz, Germany
5. Centre for Quantum Spintronics, Norwegian University of Science and Technology NTNU, 7491, Trondheim, Norway

**The spin of the electron has been a key enabler to realize spintronics devices that harness the spin degree of freedom beyond conventional charge-based electronics. In addition to spin, electrons have another degree of freedom associated with the angular momentum: orbital angular momentum. While spintronics has evolved fast and spintronics concepts are now used in magnetic sensors and nonvolatile memories, exploiting orbital angular momentum has just started to be studied. Phenomena mediated by orbital angular momentum have led to a new branch of spintronics called orbitronics. This Perspective discusses how orbitronics can constitute the next step in the evolution of spintronics by reviewing the current understandings and experimental results of orbital effects and summarizing the open questions. Prospects and challenges for devices in particular in non-volatile memory are discussed as an applications perspective.**

Spintronics has advanced information and communications technology by incorporating the spin degree of freedom of electrons into conventional electronics, which traditionally relies solely on the charge degree of freedom, enabling the electrical detection and manipulation of magnetization. Notable spintronic phenomena include the giant and tunneling magnetoresistive effects[1] and spin-transfer torque effects[2]. Discovery of these effects has led to practical applications such as magnetic sensors and magnetoresistive random access memories (MRAMs). In particular, MRAM using spin-transfer torque-induced magnetization switching for the write operation has been commercialized holding great potential to make large-scale integrated circuits more energy-efficient and underpin the future growth of the internet-of-things and artificial intelligence technologies.

Spin-orbitronics emerged as the next step beyond spintronics by incorporating the orbital degree of freedom to control electron spin via spin-orbit coupling. Unlike conventional spintronics, where magnetic materials are required to generate spin-polarized electrons, spin-orbit coupling enables the generation of spin currents in nonmagnetic systems and non-equilibrium spin accumulation at interfaces; these phenomena are known as the spin Hall effect and Rashba-Edelstein effect, respectively [see Fig. 1(a), (b) for a schematic]. As a consequence, in magnetic systems with broken spatial inversion symmetry, a magnetic torque, termed spin-orbit torque, arises when an electric current is applied, providing an alternative scheme of electrically manipulating magnetization[3,4], distinct from the spin-transfer torque. Furthermore, the spin-orbit torques enables the electrical control of the Néel vector in antiferromagnets, giving rise to the paradigm of antiferromagnetic spintronics[5]. In terms of applications, MRAMs utilizing spin-orbit torque have been developed to achieve higher-speed memories in integrated circuits[6,7], which cannot be realized using conventional spin-transfer torque MRAMs.

The most recent development termed orbitronics has attracted attention, focusing on the active control of the orbital degrees of freedom itself. In textbook knowledge, spin angular momentum and orbital angular momentum are two distinct degrees of freedom representing the angular momentum of an electron's state. While the orbital aspect has received less attention compared to spin in the context of spin-orbitronics, despite several theoretical predictions of significant orbital effects[8,9], the situation has changed recently. Firstly, in experiments on spin-orbit torques, some surprising results have been reported that cannot be explained within the framework of conventional spin-orbitronics. Secondly, effects of orbital angular momentum transport in nonequilibrium conditions, *i.e.*, orbital Hall effect and orbital Rashba-Edelstein effect as the counterparts of the spin Hall effect and Rashba-Edelstein effect, have been theoretically expanded, suggesting that these could be as important as that of spin[10]. Thirdly, the development of spin-orbit torque MRAM has highlighted the need to improve the torque generation efficiency, prompting a search for mechanisms that enhance torques.

This Perspective provides an overview of orbitronics, highlighting opportunities and challenges in surpassing conventional spintronics and spin-orbitronics. In the following sections, we review key experimental and theoretical results that go beyond spin-orbitronics, requiring one to invoke the concept of orbitronics as well as summarizing the open questions and future fundamental challenges that lie ahead. These are followed by a discussion from an applications' standpoint, focusing on the opportunities and challenges orbitronics presents in enhancing the efficiency of spin-orbit torque MRAM. Due to space limitations, we do not aim at a comprehensive review, and for further information, we refer readers to recent review articles[11,12].

**Experimental progress in orbitronics**

Experimentally, a broad range of promising results have been reported on orbitronics and orbital angular momentum in general in the past few years. When currents are injected, a wavevector-dependent orbital angular momentum of electrons can occur as will be discussed in the theory section below. A large number of predictions deal with the importance of orbital angular momentum and torques, including predictions of the orbital Hall effect[9,10] and the orbital Rashba-Edelstein effect[10,13,14], for which the mechanisms are shown schematically in Fig. 1(c),(d).

There are different experimental approaches to study and understand orbitronic effects. The most used approach is to exploit the generated torques, called orbital torque, using conventional techniques established for spin-orbit torque such as spin-torque ferromagnetic resonance or second harmonic analysis[3] in a heterostructure with a magnetic layer as shown in Fig. 2. To exploit torque effects, it is not enough to generate orbital angular momentum currents or accumulation because they do not act directly on the magnetization since in equilibrium the magnetization of most conventional magnets is based on spin. Accordingly, one needs to additionally convert the orbital angular momentum to spin angular momentum by spin-orbit coupling effects. As shown in Fig. 2(a), this can occur in the ferromagnet itself. However, not all ferromagnets are efficient at this conversion; for instance, in Refs. [15-17], it was reported that permalloy (Ni-Fe) and Ni are much more efficient than for instance CoFeB. We note that the strong dependence of the torques on the ferromagnetic material is a fingerprint used experimentally to identify orbital effects, as conventional spin currents are fully absorbed by ferromagnets within a few nm. To make all ferromagnets useful, one can introduce an additional conversion layer that converts orbital to spin angular momentum [Fig. 2(b)]. As such a conversion layer, Pt[15,18] or 4$f$ transition metals[16] have been previously used and one can also alloy these with the magnetic material. While many if not most of the experiments on orbitronics have used torque measurements, a quantitative analysis is challenging because of the number of effects involved. The generation of orbital angular momentum, the transport of orbital angular momentum in particular across interfaces (which is still not very clear[19]), the conversion to spin, the spin transport and the final transfer to the magnetic moment exerting a torque can all contribute and cannot be easily disentangled.

Beyond measuring torques, we note that such multilayers also lend themselves to studying magnetoresistance effects that can be used as the second approach to explore orbitronics[20]. The inverse effect of the torques, namely pumping of orbital angular momentum that is converted to charge currents and can also be used for THz emission, can be used as a third approach[21,22]. A fourth more direct approach is to directly image the orbital angular momentum accumulation based on circular dichroism as recently demonstrated by magneto-optical[23,24] and electron microscopy/energy-loss spectroscopy[17] methods. A fifth method is to employ orbital Hanle magnetoresistance[25], which however needs to be carefully delineated from ordinary magnetoresistance. While these different approaches have been used in the past in seminal contributions to the field, still significant further work is needed for unambiguous quantitative determination of orbitronic effects.

Previous experimental work started with early findings that oxide layers, such as $CuO_x$, enhance the torques. While this was initially not interpreted in terms of orbital effects, more recently, $CuO_x$ has been identified to be a strong source of orbital angular momentum[20,26,27]: a significant enhancement of spin torques in TmIG/Pt/$CuO_x$ heterostructures due to the orbital currents generated by the $CuO_x$ was reported. These orbital currents generated by the orbital Hall effect are efficiently converted to spin currents in the Pt leading to a sixteen-fold increase in the torques acting on the magnetic layer[26]. This key observation was complemented by the identification of the orbital Rashba-Edelstein effect in Ni-Fe/CuO$_x$ heterostructures where strong non-equilibrium orbital accumulation was converted in the Ni-Fe to a spin current that leads to strong torques on the Ni-Fe[20,22]. Beyond CuO$_x$ with a complex structure that is often difficult to control, large orbital Hall effect have been predicted for transition metals such as Nb, Ta and Ru.

By analyzing the thickness dependence, orbital current transport in ferromagnets over longer distances compared to the very short transport distance typically observed for spin currents could be identified[15,16,22,23,28]. And by the sign of the observed torques the orbital nature could be identified[15]. The conversion between orbital and spin angular momentum was systematically studied[16] and recently rare earth materials were identified to be efficient at this[29]. Enhancing torque efficiency also can be achieved by tuning the crystal symmetry of the ferromagnet[30,31]. In FeMn/CoPt bilayers, the torque efficiency varies with ferromagnet thickness depending on CoPt's crystal symmetry, attributed to orbital-dependent energy splittings and hybridizations[31]. Furthermore, recent studies have extended the magnetic materials in which the orbital torque is observed in antiferromagnets[32] or two-dimensional[33] and perpendicular[34] ferromagnets, being expected to further advance the understanding.

The reciprocal orbital pumping was also reported, corroborating the long transport length scales and the strong dependence on the ferromagnet[21]. While fundamentally one might expect that orbital to charge and charge to orbital effects could be fully reciprocal, this is far from clear with some results on local and non-local transport showing reciprocity[35] while others do not[36,37], opening further avenues for exciting research.

Beyond measuring the torques, magnetoresistance and pumping effects that rely on heterostructures with magnets, a more direct way to quantify the orbital currents generated is to image the orbital accumulation as recently pioneered in thin films using the magneto-optical Kerr effect methods[23,24]. While the Kerr effect is sensitive to both orbital and spin angular momentum, an even more direct way to identify the orbital contribution is electron magnetic linear dichroism[17] and these measurements also allow for the analysis of the orbital angular momentum diffusion length.

Beyond conventional multilayer heterostructures, recently two-dimensional van der Waals systems have been explored. The orbital Hall effect in transition metal dichalcogenides was theoretically studied intensively. In particular, while previously the transport of orbital angular momentum was interpreted in terms of the valley Hall effect, it was realized that the orbital Hall effect might play a key role. The transport of orbital angular momentum in terms of the valley Hall effect assumes that also transport of magnetic moments is generated[38]. More recent analysis has revealed that the orbital Hall effect is a more appropriate description of the transport of orbital angular momentum in bilayer transition metal dichalcogenides, as previous work based on the valley Hall effect has encountered conceptual difficulties such as the artificial separation of the Brillouin zone. The orbital Hall effect and valley Hall effect were disentangled theoretically[39], and it was predicted that the orbital Hall effect in the insulating gap even allows to attribute a Chern number confirming a possible topological nature of orbital Hall insulators[40,41] that were recently explored in rare-earth dichalcogenides[42]. Finally, it is worth pointing out that intrinsic orbital torque in single materials can also be present in two-dimensional layered compounds[43].

**Theoretical understanding for orbitronics and open questions**

The basic principle, which is described in Box 1, demonstrates that the orbital Hall effect is a universal phenomenon in all conducting solids. In addition to the bulk orbital Hall effect, the interfacial orbital Rashba-Edelstein effect also generates orbital angular momentum currents. Similar to the orbital Hall effect, the orbital Rashba-Edelstein effect exists in general even in the absence of spin-orbit coupling as it arises from orbital Rashba coupling[13]. When an orbital angular momentum current, produced by either orbital Hall effect or orbital Rashba-Edelstein effect, is injected into a ferromagnet, it exerts a magnetic torque, i.e., orbital torque[44] (more precisely, orbital angular momentum torque). This orbital angular momentum torque has become an increasingly active area of research, with the expectation that it could significantly enhance the overall efficiency of spin-orbit torque.

The usefulness of orbital degrees of freedom in spin-orbitronics depends on two closely related factors: the magnitude of orbital angular momentum torque and the relaxation process of orbital angular momentum current. Currently, the theoretical understanding of these factors is far from complete. At the same time, they represent important research topics to deepen our understanding on orbital transport in solids. Below, we outline theoretical challenges associated with these factors, which remain open questions at present.

The magnitude of orbital angular momentum torque is determined by the orbital angular momentum accumulation at the ferromagnet/normal metal interface or within the ferromagnet, in a way similar to how the spin torque magnitude is determined. Consequently, the orbital angular momentum torque depends not only on the orbital angular momentum Hall conductivity and the orbital Rashba-Edelstein effect, but also on the orbital angular momentum relaxation processes. Theoretical calculations predict large orbital angular momentum Hall conductivities in various materials[45,46], driven by the orbital-lattice coupling described by the equations above. This coupling, mediated by crystal fields with energy scales on the order of 1 eV, is exceptionally strong. Given that this coupling efficiently generates orbital angular momentum currents, it must also dissipate them efficiently, as the coupling itself cannot create orbital angular momentum but only facilitates its exchange between two reservoirs (i.e., orbital and lattice). The latter strong dissipation of orbital angular momentum is in line with the well-known phenomenon of orbital quenching in solids. As a result, significant non-conservation of orbital angular momentum currents is expected in solids.

Regarding the orbital angular momentum relaxation, several experiments have reported a long orbital relaxation length[15,16,22,23,28], much longer than the spin relaxation length, which is challenging to reconcile theoretically. As mentioned above, the crystal field coupling ($\sim 1$ eV), which governs orbital relaxation, is much stronger than the spin-orbit coupling ($10 \sim 100$ meV), which is responsible for spin relaxation. In this context, these experimental observations of long-range orbital transport call for further theoretical development. Related to this issue, there are two conflicting theories. One first-principles calculation proposes long-range orbital torques in ferromagnets, attributing it to the physics of hotspots, i.e., $k$-points with nearly degenerate orbital bands[47]. In contrast, another first-principles calculation predicts an extremely short orbital diffusion length in normal metals[19]. Since there is no clear reason why orbital relaxation should differ substantially between ferromagnets and normal metals, this discrepancy leaves the orbital relaxation process an unresolved issue. From a practical perspective, moreover, a shorter orbital relaxation length is preferable, as torque efficiency is proportional to $[1 - \exp(-d/\lambda)]$, where $d$ and $\lambda$ are the thickness and the orbital diffusion length of normal metal, respectively.

In summary, the long orbital relaxation length remains an intriguing yet unresolved issue, with conflicting evidence from experimental observations and theoretical calculations. Understanding the mechanisms behind orbital relaxation is one crucial point for advancing our knowledge of orbital transport in the presence of strong crystal field coupling.

## Opportunities and challenges for applications

Following an overview of fundamental research from spin-orbitronics to orbitronics, we now discuss the opportunities and challenges orbitronics presents for practical applications. Like the spin-orbit torque, the orbital torque has several potential applications, including MRAM, oscillators, random number generators, logic devices, and THz emitters. In this section, we focus primarily on the application to MRAM devices and discuss the issues faced by conventional spin-orbit torque, the advantages the orbital torque could provide, and the challenges the orbital torque should overcome.

Box 2 illustrates the key requirements for MRAM applications, along with a schematic of a three-terminal magnetic tunnel junction device array, where spin-orbit torques and orbital torques can be used for the write operation. W/CoFeB/MgO system is known to exhibit large current-induced torque due to the significant spin Hall effect of W[48], reasonably meeting these requirements and being employed in spin-orbit torque MRAM demonstrations[6,7]. However, this type of material system with a relatively large resistivity has an intrinsic trade-off between low write current and low channel resistance, prompting the development and engineering of alternative material systems as well as the exploration of alternative driving mechanisms that achieve higher torque generation efficiency with lower resistance. Various unconventional physical mechanisms for current-induced torque generation have been extensively studied in material systems beyond the spin Hall effect in $5d$ heavy metals such as W, but it remains challenging to satisfy all these requirements simultaneously.

Given these circumstances, the orbital torque could emerge as a potential alternative driving mechanism owing to several favorable characteristics. For instance, as described earlier, the generation of flow and accumulation of orbital angular momentum does not rely on the spin-orbit coupling unlike the spin. This allows for using light metals such as Ti, V, and Cr, which typically exhibit much smaller resistivity than W. In addition, these materials are generally compatible with the integration process, potentially fulfilling the requirements.

However, the orbital torque also presents some specific challenges. The first challenge is increasing the magnitude of generated torque relative to the current density, often quantified by a dimensionless parameter $\xi = 2eH_{\text{eff}}M_S t/\hbar J$ ($e$: elementary charge, $H_{\text{eff}}$: effective field of anti-damping torque, $M_S$: saturation magnetization, $t$: thickness of ferromagnetic layer, $\hbar$: Dirac constant, and $J$: current density), which corresponds to the effective spin Hall angle when the torque originates from the spin Hall effect. To date, the reported $\xi$ of the orbital torque is up to around 0.3, but much higher values are required to achieve low write current and high endurance. The second challenge is the generation of a significant orbital torque in systems that are compatible with high tunneling magnetoresistance. Although considerable orbital torques have been observed with ferromagnetic materials with large spin-orbit coupling, such as Ni and Ni-Fe[15,49] or in stacks including high spin-orbit coupling nonmagnetic layers like Pt[15,18] and Tb[16] as described earlier, these systems are generally incompatible with high tunneling magnetoresistance. Achieving substantial orbital torque in bcc ferromagnets such as CoFeB is desirable so that coherent tunneling of MgO barrier is harnessed. High spin-orbit ferromagnets in general also have large damping constant and often require special seed layer, posing a challenge for applications. Third, while the experimental studies reported considerably large thicknesses of both the nonmagnetic and ferromagnetic layers at which large orbital torque emerges as discussed earlier, these thicknesses are necessary to be reduced by shortening the orbital angular momentum diffusion length to meet the low write-current requirement. Material engineering or the development of new material systems to address these challenges will be crucial for the successful implementation of MRAMs utilizing the orbital torque.

As developments towards MRAM applications, several works have been carried out. For instance, orbital torque-induced switching properties in devices equipping a 100-nm ferromagnetic dot with a perpendicular easy axis was studied, showing 50% reduction of switching power compared to spin-orbit torque devices with Pt[50]. Meanwhile, an apples-to-apples comparison between archetypal spin-orbit torque- and orbital torque-dominant systems (W/CoFeB/MgO and Cr/CoFeB/MgO, respectively) was performed, clarifying the opportunities and challenges of devices harnessing the orbital torque[51]. Beyond memory, large angular momentum generation has been exploited for THz emission[22,52] and applications in unconventional computing[34] can also benefit from orbitronics highlighting the broad impact of this field.

## Outlook

The orbitronics is attractive as an orbital texture exists even in the absence of spin-orbit coupling, allowing for using light, abundant, cheap, and environmentally non-hazardous materials. As described above, while solid evidence for the importance of orbital effects has been established both experimentally and theoretically, the individual processes occurring during their generation are not well understood. This includes strong materials-dependence of the generation of orbital angular momentum, the transport of orbital angular momentum over long distances and across interfaces between different materials, the conversion of the orbital angular momentum to spin angular momentum, and the orders of magnitude enhanced torques in particular expected for two-dimensional materials. Meanwhile, the sensitivity of orbital angular momentum to the chemical and crystal environments suggests the possibility to tune the effects by various means such as crystal and interface engineering. Also, the orbital angular momentum is not dissipated by Gilbert damping in the same manner as the spin and is expected to be much less sensitive to magnetic perturbations, making it potentially more robust. Finally, orbital angular momentum of electrons can be potentially coupled with that of various quasi-particles such as photon and phonon, showing promise to create new branches of condensed-matter physics by fusing with photonics, phononics and others.

## Acknowledgements
S.F. acknowledges the financial support from MEXT X-NICS (JPJ011438), JSPS Kakenhi (24H00039 and 24H02235), JST-ASPIRE JPMJAP2322, and RIEC Cooperative Research Projects. K.J.L. acknowledges financial supports from the National Research Foundation of Korea (2022M3H4A1A04096339, RS-2024-00410027) and Samsung Electronics (IO201019-07699-01). M.K. acknowledges TopDyn, the German Research Foundation (TRR 173-268565370 Spin+X: A01, A11, B02 and project 358671374), the Horizon Europe Framework (Grant No. 101070290 (NIMFEIA), Grant No. 101070287 (Swan-on-Chip)) and Grant No. 101129641 (OBELIX), the European Research Council (Grant No. 856538 (3D MAGiC)), and the Research Council of Norway through its Centers of Excellence funding scheme, Project No. 262633 "QuSpin."

**BOX 1 Orbital angular momentum generation from momentum-space orbital texture**

The orbital Hall effect arises from momentum-space orbital texture (in short, orbital texture)[8,10], where the orbital refers to a linear combination of real orbitals, which do not carry orbital angular momentum. These real orbitals are described by operators of even-order symmetrized products of the orbital angular momentum operator $L$[8,53]. For *p*-orbitals, for instance, the operators are described by $\{L_i, L_j\}$, the anticommutation of $L_i$ and $L_j$.

Importantly, orbital textures exist even in systems that preserve both time-reversal and inversion symmetries, making its presence general. For such systems, the Hamiltonian for *p*-orbitals is[8]:

$$\mathcal{H} = Ak^2 + \sum_{i,j=x,y,z} B_{ij}\{k_i, k_j\}\{L_i, L_j\}, \tag{1}$$

where $A$, $B_{ij}$ are material constants and $k$ is the crystal momentum. Assuming $B_{ij} = B$, which is a reasonable approximation near the $\Gamma$ point, the second term of Eq. (1) simplifies to $B(\mathbf{L} \cdot \mathbf{k})^2$ [Ref. [8]], clearly illustrating the momentum-dependent texture of $\mathbf{L}^2$, i.e., the orbital texture.

When an electric field is applied, it induces the dynamics of $\{L_i, L_j\}$ [Ref. [53]]:

$$\frac{d\{L_i, L_j\}}{dt} = \frac{[\{L_i, L_j\}, \mathcal{H}]}{i\hbar} = \sum_k C_{ijk} L_k, \tag{2}$$

creating $L_k$ (orbital angular momentum), which is the underlying origin of the orbital Hall effect.

**BOX 2 Requirements on the three-terminal magnetic tunnel junction device for nonvolatile-memory applications.**

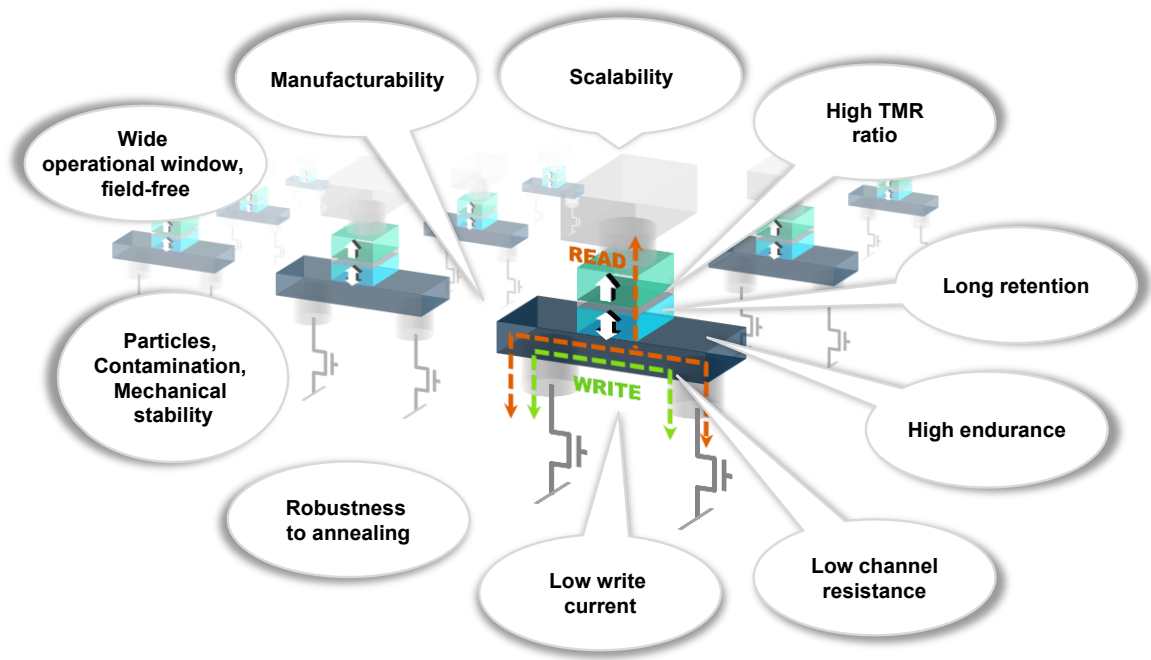

The three terminal magnetic tunnel junction devices illustrated here have separated write and read current paths (shown by broken arrows), allowing for higher speed operation than the two-terminal counterpart used in spin-transfer torque-MRAMs. Several representative requirements satisfied simultaneously for practical applications are indicated by balloons in the schematic. The high tunneling magnetoresistance (TMR) ratio necessary for stable read operation requires a material system compatible with the coherent tunneling such as CoFeB/MgO. The long retention for stable data storage requires sufficiently high magnetic anisotropy energy of the free layer. Materials with long retention capability are also beneficial for the write operation with low error rate. The high endurance, necessary to replace semiconductor working memories, requires a channel material having sufficient durability against electromigration. Low channel resistance is necessary for low-voltage operations, ideally lower than about 0.1 V. The low write current, roughly lower than 100 µA necessary for low-power operation and small cell (transistor) size, requires a material system generating large torque with respect to the applied current. The robustness to annealing is necessary for formation of interconnects and process of packaging where high-temperature annealing typically at 400°C is conducted. Negligible particles and contamination, as well as high mechanical stability, are a general requirement for the integration process. The wide operational window and field-free operation represents that the device functions in a wide environmental range in terms of temperature, magnetic field, etc. The widest temperature range required for automobile applications is -40 – 150°C (Grade 0 of AEC Q100). Magnetic field-free operation is of course necessary. The manufacturability represents that the stack can be formed on amorphous substrates such as Si-O and Si-N, stack can be easily etched with sufficient process margin, and the properties do not significantly change with patterning and encapsulation processes. The scalability represents that the favorable properties are maintained or improved with the miniaturization of devices in order to be integrated into the cutting-edge CMOS circuits.

**References**

1 Hirohata, A. *et al.* Review on spintronics: Principles and device applications. *Journal of Magnetism and Magnetic Materials* **509**, 166711, doi:https://doi.org/10.1016/j.jmmm.2020.166711 (2020).
2 Brataas, A., Kent, A. D. & Ohno, H. Current-induced torques in magnetic materials. *Nat. Mater.* **11**, 372-381, doi:10.1038/nmat3311 (2012).
3 Manchon, A. *et al.* Current-induced spin-orbit torques in ferromagnetic and antiferromagnetic systems. *Rev. Mod. Phys.* **91**, doi:10.1103/RevModPhys.91.035004 (2019).
4 Shao, Q. *et al.* Roadmap of Spin–Orbit Torques. *IEEE Trans. Magn.* **57**, 1-39, doi:10.1109/tmag.2021.3078583 (2021).
5 Jungwirth, T., Marti, X., Wadley, P. & Wunderlich, J. Antiferromagnetic spintronics. *Nat. Nanotechnol.* **11**, 231-241, doi:10.1038/nnano.2016.18 (2016).
6 Natsui, M. *et al.* Dual-Port SOT-MRAM Achieving 90-MHz Read and 60-MHz Write Operations Under Field-Assistance-Free Condition. *IEEE J. Sol.-Sta. Cir.* **56**, 1116-1128, doi:10.1109/jssc.2020.3039800 (2021).
7 Song, M. Y. *et al.* High speed (1ns) and low voltage (1.5V) demonstration of 8Kb SOT-MRAM array. in *2022 IEEE Symposium on VLSI Technology and Circuits (VLSI Technology and Circuits)*. 377-378 doi:10.1109/VLSITechnologyandCir46769.2022.9830149).
8 Bernevig, B. A., Hughes, T. L. & Zhang, S.-C. Orbitronics: The Intrinsic Orbital Current in *p*-Doped Silicon. *Phys. Rev. Lett.* **95**, 066601, doi:10.1103/PhysRevLett.95.066601 (2005).
9 Kontani, H., Tanaka, T., Hirashima, D. S., Yamada, K. & Inoue, J. Giant Orbital Hall Effect in Transition Metals: Origin of Large Spin and Anomalous Hall Effects. *Phys. Rev. Lett.* **102**, 016601, doi:10.1103/PhysRevLett.102.016601 (2009).
10 Go, D., Jo, D., Kim, C. & Lee, H.-W. Intrinsic Spin and Orbital Hall Effects from Orbital Texture. *Phys. Rev. Lett.* **121**, 086602, doi:10.1103/PhysRevLett.121.086602 (2018).
11 Go, D., Jo, D., Lee, H.-W., Kläui, M. & Mokrousov, Y. Orbitronics: Orbital currents in solids. *Europhys. Lett.* **135**, 37001, doi:10.1209/0295-5075/ac2653 (2021).
12 Jo, D., Go, D., Choi, G.-M. & Lee, H.-W. Spintronics meets orbitronics: Emergence of orbital angular momentum in solids. *npj Spintronics* **2**, doi:10.1038/s44306-024-00023-6 (2024).
13 Park, S. R., Kim, C. H., Yu, J., Han, J. H. & Kim, C. Orbital-Angular-Momentum Based Origin of Rashba-Type Surface Band Splitting. *Phys. Rev. Lett.* **107**, 156803, doi:10.1103/PhysRevLett.107.156803 (2011).
14 Kim, J. *et al.* Nontrivial torque generation by orbital angular momentum injection in ferromagnetic-metal/Cu/$Al_2O_3$ trilayers. *Physical Review B* **103**, L020407, doi:10.1103/PhysRevB.103.L020407 (2021).
15 Bose, A. *et al.* Detection of long-range orbital-Hall torques. *Physical Review B* **107**, 134423, doi:10.1103/PhysRevB.107.134423 (2023).
16 Sala, G. & Gambardella, P. Giant orbital Hall effect and orbital-to-spin conversion in 3*d*, 5*d*, and 4*f* metallic heterostructures. *Phys. Rev. Res.* **4**, 033037, doi:10.1103/PhysRevResearch.4.033037 (2022).
17 Idrobo, J. C. *et al.* Direct observation of nanometer-scale orbital angular momentum accumulation. arXiv:2403.09269 (2024). <https://arxiv.org/abs/2403.09269>.
18 Lee, S. *et al.* Efficient conversion of orbital Hall current to spin current for spin-orbit torque switching. *Commun. Phys.* **4**, 234, doi:10.1038/s42005-021-00737-7 (2021).
19 Rang, M. & Kelly, P. J. Orbital relaxation length from first-principles scattering calculations. *Physical Review B* **109**, 214427, doi:10.1103/PhysRevB.109.214427 (2024).
20 Ding, S. *et al.* Observation of the Orbital Rashba-Edelstein Magnetoresistance. *Phys. Rev. Lett.* **128**, 067201, doi:10.1103/PhysRevLett.128.067201 (2022).
21 Hayashi, H., Go, D., Haku, S., Mokrousov, Y. & Ando, K. Observation of orbital pumping. *Nature Electronics* **7**, 646-652, doi:10.1038/s41928-024-01193-1 (2024).
22 Seifert, T. S. *et al.* Time-domain observation of ballistic orbital-angular-momentum currents with giant relaxation length in tungsten. *Nat. Nanotechnol.* **18**, 1132-1138, doi:10.1038/s41565-023-01470-8 (2023).
23 Choi, Y.-G. *et al.* Observation of the orbital Hall effect in a light metal Ti. *Nature* **619**, 52-56, doi:10.1038/s41586-023-06101-9 (2023).
24 Lyalin, I., Alikhah, S., Berritta, M., Oppeneer, P. M. & Kawakami, R. K. Magneto-Optical Detection of the Orbital Hall Effect in Chromium. *Phys. Rev. Lett.* **131**, 156702, doi:10.1103/PhysRevLett.131.156702 (2023).
25 Sala, G., Wang, H., Legrand, W. & Gambardella, P. Orbital Hanle Magnetoresistance in a 3*d* Transition Metal. *Phys. Rev. Lett.* **131**, 156703, doi:10.1103/PhysRevLett.131.156703 (2023).
26 Ding, S. *et al.* Harnessing Orbital-to-Spin Conversion of Interfacial Orbital Currents for Efficient Spin-Orbit Torques. *Phys. Rev. Lett.* **125**, 177201, doi:10.1103/PhysRevLett.125.177201 (2020).
27 Kim, J. *et al.* Oxide layer dependent orbital torque efficiency in ferromagnet/Cu/oxide heterostructures. *Physical Review Materials* **7**, L111401, doi:10.1103/PhysRevMaterials.7.L111401 (2023).
28 Hayashi, H. *et al.* Observation of long-range orbital transport and giant orbital torque. *Commun. Phys.* **6**, 32, doi:10.1038/s42005-023-01139-7 (2023).
29 Ding, S., Kang, M.-G., Legrand, W. & Gambardella, P. Orbital Torque in Rare-Earth Transition-Metal Ferrimagnets. *Phys. Rev. Lett.* **132**, 236702, doi:10.1103/PhysRevLett.132.236702 (2024).
30 Fukunaga, R., Haku, S., Gao, T., Hayashi, H. & Ando, K. Impact of crystallinity on orbital torque generation in ferromagnets. *Physical Review B* **109**, 144412, doi:10.1103/PhysRevB.109.144412 (2024).
31 Gao, T. *et al.* Control of dynamic orbital response in ferromagnets via crystal symmetry. *Nature Physics* **20**, 1896-1903, doi:10.1038/s41567-024-02648-0 (2024).
32 Zheng, Z. *et al.* Effective electrical manipulation of a topological antiferromagnet by orbital torques. *Nat. Commun.* **15**, 745, doi:10.1038/s41467-024-45109-1 (2024).
33 Li, D. *et al.* Room-temperature van der Waals magnetoresistive memories with data writing by orbital current in the Weyl semimetal $TaIrTe_4$. *Physical Review B* **110**, 035423, doi:10.1103/PhysRevB.110.035423 (2024).
34 Zhao, Q. *et al.* Efficient Charge-to-Orbit Current Conversion for Orbital Torque Based Artificial Neurons and Synapses. *Advanced Electronic Materials* **n/a**, 2400721, doi:https://doi.org/10.1002/aelm.202400721.
35 Otani, Y. *et al.* Nonlocal Electrical Detection of Reciprocal Orbital Edelstein Effect. (2024). <https://doi.org/10.21203/rs.3.rs-5107751/v1>.
36 Mendoza-Rodarte, J. A., Cosset-Chéneau, M., van Wees, B. J. & Guimarães, M. H. D. Efficient Magnon Injection and Detection via the Orbital Rashba-Edelstein Effect. *Phys. Rev. Lett.* **132**, 226704, doi:10.1103/PhysRevLett.132.226704

37 Ledesma-Martin, J. O. *et al.* Nonreciprocity in Magnon Mediated Charge-Spin-Orbital Current Interconversion. *Nano Letters* **25**, 3247-3252, doi:10.1021/acs.nanolett.4c06056 (2025).
38 Xiao, D., Liu, G.-B., Feng, W., Xu, X. & Yao, W. Coupled Spin and Valley Physics in Monolayers of $MoS_2$ and Other Group-VI Dichalcogenides. *Phys. Rev. Lett.* **108**, 196802, doi:10.1103/PhysRevLett.108.196802 (2012).
39 Cysne, T. P. *et al.* Disentangling Orbital and Valley Hall Effects in Bilayers of Transition Metal Dichalcogenides. *Phys. Rev. Lett.* **126**, 056601, doi:10.1103/PhysRevLett.126.056601 (2021).
40 Costa, M. *et al.* Connecting Higher-Order Topology with the Orbital Hall Effect in Monolayers of Transition Metal Dichalcogenides. *Phys. Rev. Lett.* **130**, 116204, doi:10.1103/PhysRevLett.130.116204 (2023).
41 Ji, S., Quan, C., Yao, R., Yang, J. & Li, X. a. Reversal of orbital Hall conductivity and emergence of tunable topological quantum states in orbital Hall insulators. *Physical Review B* **109**, 155407, doi:10.1103/PhysRevB.109.155407 (2024).
42 Zeer, M. *et al.* Spin and orbital transport in rare-earth dichalcogenides: The case of $EuS_2$. *Physical Review Materials* **6**, 074004, doi:10.1103/PhysRevMaterials.6.074004 (2022).
43 Saunderson, T. G., Go, D., Blügel, S., Kläui, M. & Mokrousov, Y. Hidden interplay of current-induced spin and orbital torques in bulk $Fe_3GeTe_2$. *Phys. Rev. Res.* **4**, L042022, doi:10.1103/PhysRevResearch.4.L042022 (2022).
44 Go, D. & Lee, H.-W. Orbital torque: Torque generation by orbital current injection. *Phys. Rev. Res.* **2**, 013177, doi:10.1103/PhysRevResearch.2.013177 (2020).
45 Jo, D., Go, D. & Lee, H.-W. Gigantic intrinsic orbital Hall effects in weakly spin-orbit coupled metals. *Physical Review B* **98**, 214405, doi:10.1103/PhysRevB.98.214405 (2018).
46 Salemi, L. & Oppeneer, P. M. First-principles theory of intrinsic spin and orbital Hall and Nernst effects in metallic monoatomic crystals. *Physical Review Materials* **6**, 095001, doi:10.1103/PhysRevMaterials.6.095001 (2022).
47 Go, D. *et al.* Long-Range Orbital Torque by Momentum-Space Hotspots. *Phys. Rev. Lett.* **130**, 246701, doi:10.1103/PhysRevLett.130.246701 (2023).
48 Pai, C.-F. *et al.* Spin transfer torque devices utilizing the giant spin Hall effect of tungsten. *Appl. Phys. Lett.* **101**, 122404, doi:10.1063/1.4753947 (2012).
49 Lee, D. *et al.* Orbital torque in magnetic bilayers. *Nat. Commun.* **12**, 6710, doi:10.1038/s41467-021-26650-9 (2021).
50 Gupta, R. *et al.* Harnessing orbital Hall effect in spin-orbit torque MRAM. *Nat. Commun.* **16**, 130, doi:10.1038/s41467-024-55437-x (2025).
51 Chiba, S., Marui, Y., Ohno, H. & Fukami, S. Comparative Study of Current-Induced Torque in Cr/CoFeB/MgO and W/CoFeB/MgO. *Nano Letters* **24**, 14028–14033, doi:10.1021/acs.nanolett.4c03809 (2024).
52 Liu, Y. *et al.* Efficient Orbitronic Terahertz Emission Based on CoPt Alloy. *Advanced Materials* **36**, 2404174, doi:https://doi.org/10.1002/adma.202404174 (2024).
53 Han, S., Lee, H.-W. & Kim, K.-W. Orbital Dynamics in Centrosymmetric Systems. *Phys. Rev. Lett.* **128**, 176601, doi:10.1103/PhysRevLett.128.176601 (2022).

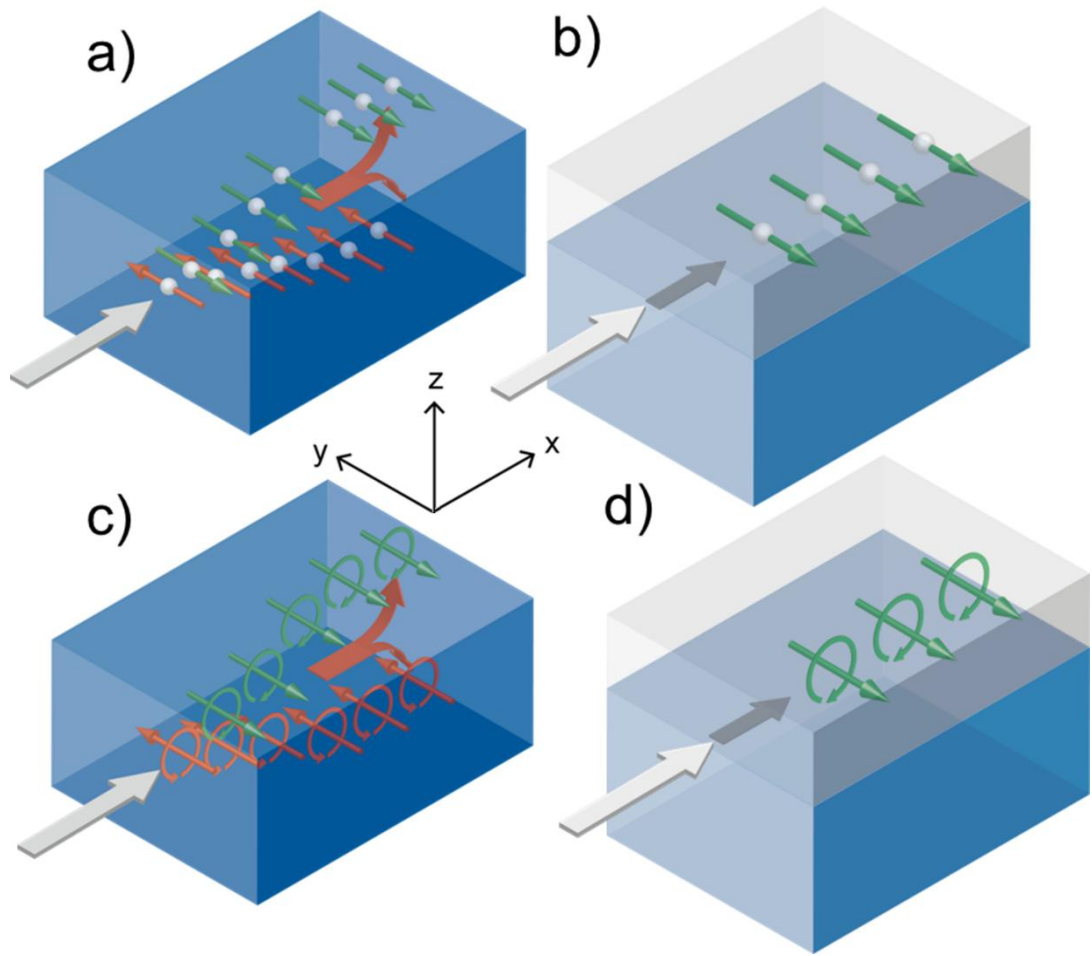

**Figure 1: Schematic depictions of the spin-orbit torque and orbital torque mechanisms.** (a) The spin Hall effect converts a charge current (grey arrow) into a transverse spin current and (c) the equivalent process of the orbital Hall effect generates a transverse orbital angular momentum current. The corresponding mechanism to (b) the Rashba-Edelstein effect that converts a charge current into a non-equilibrium spin density is (d) the orbital Rashba-Edelstein effect where a non-equilibrium orbital angular momentum density is generated at the interface between a non-magnetic material (blue) and for instance a ferromagnet (grey)

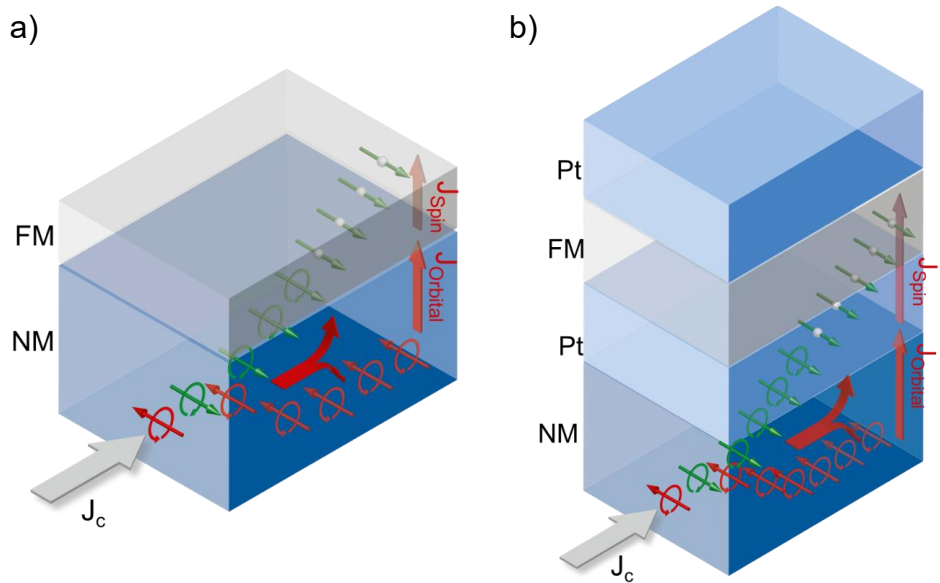

**Figure 2: Schematic of the concept of orbital torques in a multilayer with a non-magnetic material NM generating an orbital current**. a) The transverse orbital current $J_{Orbital}$ diffuses into the ferromagnet and is converted to a spin current $J_{Spin}$ that acts as a torque on the magnetization in the ferromagnet. The material-dependent efficiency of the $J_{Orbital}\rightarrow J_{Spin}$ conversion underlies the strong dependence of the torques on the ferromagnetic material, which is a signature of orbital torques. To enhance the efficiency of the orbital to spin current conversion, an additional strong spin-orbit coupling material, such as Pt can be inserted as a conversion layer. This enables the conversion $J_{Orbital} \rightarrow J_{Spin}$. A second Pt layer can be added on top to compensate for spin current generated by the spin Hall effect in the Pt layer.